\documentclass[sigconf]{acmart}
\renewcommand\footnotetextcopyrightpermission[1]{}
\def\BibTeX{{\rm B\kern-.05em{\sc i\kern-.025em b}\kern-.08emT\kern-.1667em\lower.7ex\hbox{E}\kern-.125emX}}

\setcopyright{none}

\usepackage{xspace}
\usepackage{listings}
\usepackage{tikz}
\usepackage[noend]{algpseudocode}
\usepackage{tabularx,booktabs}
\usepackage[english]{babel}
\usepackage{pdfpages}
\usepackage{adjustbox}
\usepackage[bookmarks=false]{hyperref}
\usepackage{enumitem}

\let\oldReturn\Return
\renewcommand{\Return}{\State\oldReturn}

\definecolor{darkred}{rgb}{0.75,0,0}
\definecolor{darkblue}{rgb}{0,0,0.75}
\definecolor{darkgreen}{rgb}{0,0.75,0}
\definecolor{ncs}{rgb}{0.0, 0.53, 0.74}

\newcommand{\hide}[1]{}

\newcommand{\sarif}{\textsc{Sarif}\xspace}
\newcommand{\sasp}{\textsc{Sasp}\xspace}
\newcommand{\cognicrypt}{\textsc{CogniCrypt}\xspace}
\newcommand{\crysl}{\textsc{CrySL}\xspace}

\definecolor{pblue}{rgb}{0.13,0.13,1}
\definecolor{pgreen}{rgb}{0,0.5,0}
\definecolor{pred}{rgb}{0.9,0,0}
\definecolor{pgrey}{rgb}{0.46,0.45,0.48}
\lstdefinelanguage{json}{
    stepnumber=1,
    showstringspaces=false,
    breaklines=true,
    literate=
     *{0}{{{\color{pblue}0}}}{1}
      {1}{{{\color{pblue}1}}}{1}
      {2}{{{\color{pblue}2}}}{1}
      {3}{{{\color{pblue}3}}}{1}
      {4}{{{\color{pblue}4}}}{1}
      {5}{{{\color{pblue}5}}}{1}
      {6}{{{\color{pblue}6}}}{1}
      {7}{{{\color{pblue}7}}}{1}
      {8}{{{\color{pblue}8}}}{1}
      {9}{{{\color{pblue}9}}}{1}
      {:}{{{\color{pgreen}{:}}}}{1}
      {,}{{{\color{pgreen}{,}}}}{1}
      {\{}{{{\color{pred}{\{}}}}{1}
      {\}}{{{\color{pred}{\}}}}}{1}
      {[}{{{\color{pred}{[}}}}{1}
      {]}{{{\color{pred}{]}}}}{1},
}

\algnewcommand\algorithmicswitch{\textbf{switch}}
\algnewcommand\algorithmiccase{\textbf{case}}
\algnewcommand\algorithmicassert{\texttt{assert}}
\algnewcommand\Assert[1]{\State \algorithmicassert(#1)}%
\algdef{SE}[SWITCH]{Switch}{EndSwitch}[1]{\algorithmicswitch\ #1\ \algorithmicdo}{\algorithmicend\ \algorithmicswitch}%
\algdef{SE}[CASE]{Case}{EndCase}[1]{\algorithmiccase\ #1}{\algorithmicend\ \algorithmiccase}%
\algtext*{EndSwitch}%
\algtext*{EndCase}%

\usetikzlibrary{fit,calc,trees,positioning,arrows,chains,shapes.geometric,decorations.pathreplacing,decorations.pathmorphing,shapes,matrix,shapes.symbols}
\tikzset{
>=stealth',
  punktchain/.style={
    rectangle, 
    rounded corners, 
    draw=black, very thick,
    text width=5em, 
    minimum height=3em, 
    text centered, 
    on chain},
  line/.style={draw, thick, <-},
  element/.style={
    tape,
    top color=white,
    bottom color=blue!50!black!60!,
    minimum width=5em,
    draw=blue!40!black!90, very thick,
    text width=5em, 
    minimum height=3.5em, 
    text centered, 
    on chain},
  every join/.style={->, thick,shorten >=1pt},
  tuborg/.style={decorate},
  tubnode/.style={midway, right=2pt},
}

\begin{document}

\thispagestyle{empty}

\begin{center}	

	\colorbox{ncs}{
		\begin{minipage}{17cm}
			\begin{minipage}{.74\textwidth}
  			{\color{white}
				 \vspace{1.7cm}
				{\hspace{1.1em}\fontsize{30}{60}\selectfont\textbf{Technical Report}} \\ [20pt]
				 \vspace{.2cm}{\hspace{1.1em}\huge\textbf{Paderborn University}} \\
				 \vspace{.2cm}{\hspace{1.1em}\huge\textbf{tr-ri-19-359}} \\
				 \vspace{.2cm}{\hspace{1.1em}\huge\textbf{\today}} 
				\vspace{1.5cm}
			}
		\end{minipage}%
		\begin{minipage}{.32\textwidth}
  				\includegraphics[width=4cm]{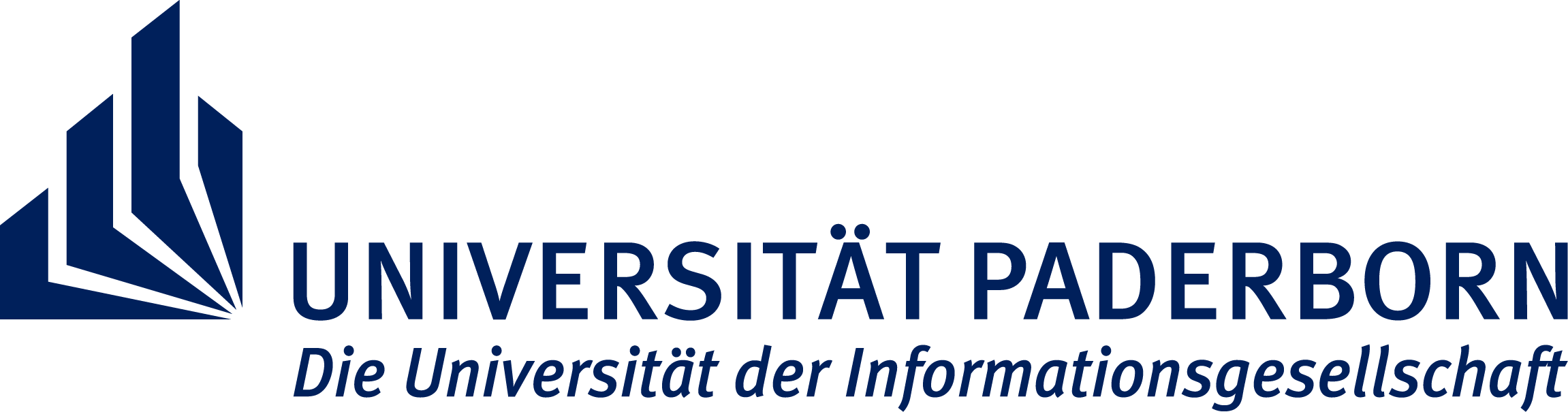} \\ [10pt]
  				\includegraphics[width=4cm]{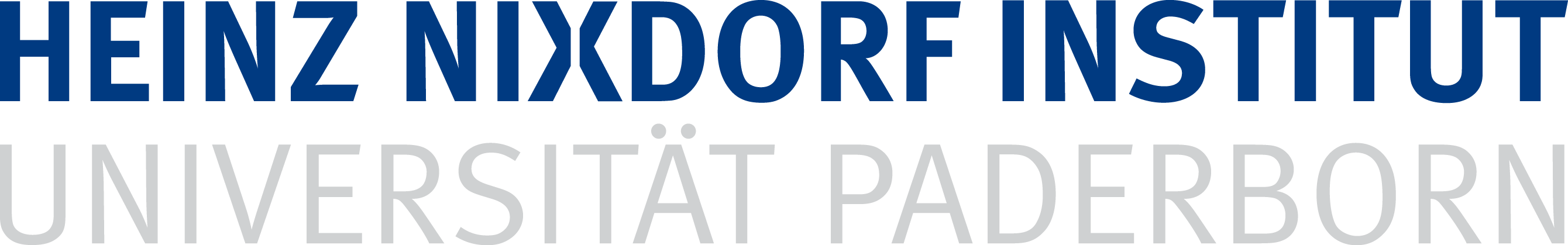} \\ [10pt]
  				\includegraphics[width=4cm]{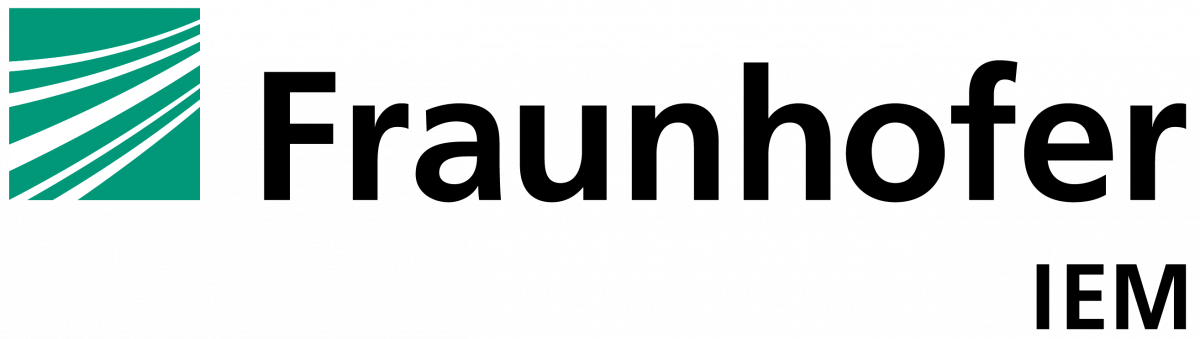}
		\end{minipage}%
		\end{minipage}
	}
		
	 \vspace{1.5cm}
	{\fontsize{20}{60}\selectfont\textbf{Integration of the Static Analysis Results}} \\[10pt]
	{\fontsize{20}{60}\selectfont\textbf{Interchange Format (SARIF) in CogniCrypt}} 
	 
	\vspace{2cm}
	\includegraphics[height=4.5cm]{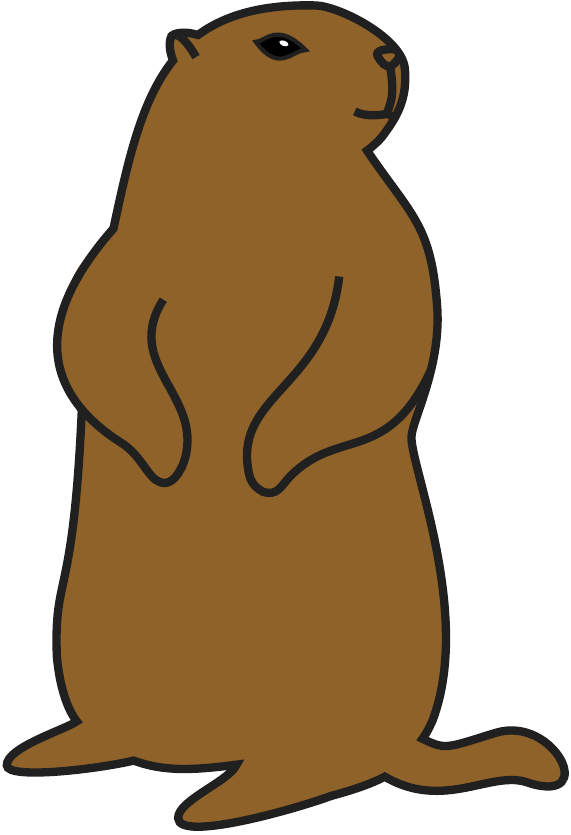}
	\vspace{2cm}
	 
	\begin{minipage}{17cm}
  		{\Large
			\textbf{Authors:} \\ [5pt]
			{Sriteja Kummita (Paderborn University)} \\
			{Goran Piskachev (Fraunhofer IEM)} \\[5pt]
		}
	\end{minipage}	
	
	\vspace{.2cm}
	\colorbox{ncs}{
		\begin{minipage}{17cm}
			{ \hspace{17cm}
			\vspace{.5cm}}
		\end{minipage}
	}

\end{center}
\newpage

\clearpage
\pagenumbering{arabic}

\title{Integration of the Static Analysis Results Interchange Format in CogniCrypt}

\author{Sriteja Kummita}
\email{sritejak@campus.uni-paderborn.de}
\affiliation{
  \institution{Paderborn University}
  \country{Germany}
}

\author{Goran Piskachev}
\email{goran.piskachev@iem.fraunhofer.de}
\affiliation{
  \institution{Fraunhofer IEM}
  \country{Germany}
}

\begin{abstract}

\textbf{Background -} Software companies increasingly rely on static analysis tools to detect potential bugs and security vulnerabilities in their software products. In the past decade, more and more commercial and open-source static analysis tools have been developed and are maintained. Each tool comes with its own reporting format, preventing an easy integration of multiple analysis tools in a single interface, such as the Static Analysis Server Protocol (\sasp). In 2017, a collaborative effort in industry, including Microsoft and GrammaTech, has proposed the Static Analysis Results Interchange Format (\sarif) to address this issue. \sarif is a standardized format in which static analysis warnings can be encoded, to allow the import and export of analysis reports between different tools.

\textbf{Purpose -} This paper explains the \sarif format through examples and presents a proof of concept of the connector that allows the static analysis tool \cognicrypt to generate and export its results in \sarif format.

\textbf{Design/Approach -} We conduct a cross-sectional study between the SARIF format and \cognicrypt's output format before detailing the implementation of the connector. The study aims to find the components of interest in \cognicrypt that the \sarif export module can complete.

\textbf{Originality/Value -} The integration of \sarif into \cognicrypt described in this paper can be reused to integrate \sarif into other static analysis tools.

\textbf{Conclusion -} After detailing the \sarif format, we present an initial implementation to integrate \sarif into \cognicrypt. After taking  advantage of all the features provided by \sarif, \cognicrypt will be able to support \sasp.
\end{abstract}

%
%



\maketitle

\sloppy

\section{Introduction}

In order to detect errors in their programs, software companies and individual developers use static analysis tools to analyze their software. From the correctness of the program, to security vulnerabilities, to compliance with given standards, to performance, static analysis is widely used in practice. Current tools typically generate reports in their own format for their own interface, or provide means to export general reports in XML or PDF format for example. As a result, software developers often experience a significant overhead parsing and aggregating the reports generated by different analysis tools in order to obtain one complete report. To address this problem, CA Technologies~\cite{catech}, Cryptsoft~\cite{cryptosoft}, FireEye~\cite{fireeye}, GrammaTech~\cite{grammaTech}, Hewlett Packard Enterprise (HPE)~\cite{hpe}, Micro Focus~\cite{microfocus}, Microsoft~\cite{microsoft}, Semmle~\cite{semmle}, and others, proposed a common reporting format for all static analysis tools, the Static Analysis Results Interchange Format, abbreviated as \sarif.

\begin{figure}[t]
	\includegraphics[width=0.48\textwidth]{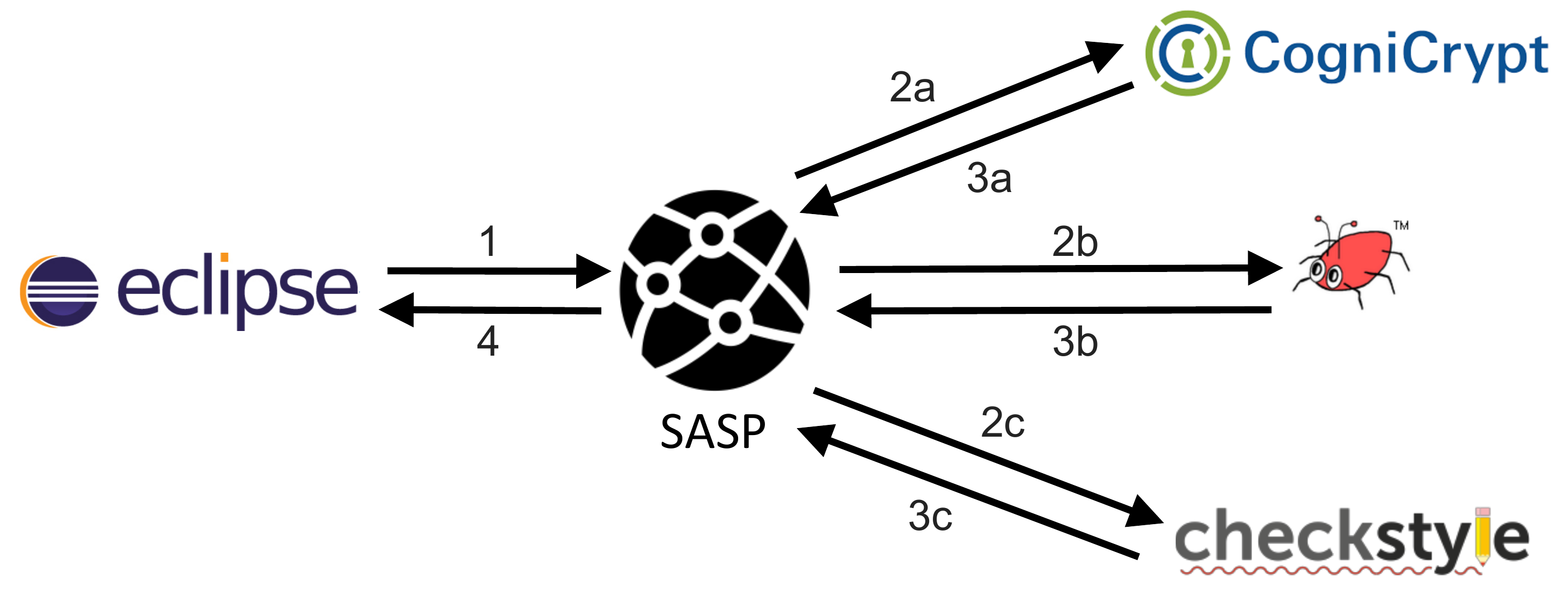}
	\caption{An illustration of \sasp integrated with static analysis clients. When a client (e.g., Eclipse), requests static analysis results from SASP (1), SASP requests results from all other clients (2a--c). It receives them (3a--c), and sends the aggregated report back to Eclipse (4).
	}
	\label{fig:sasp}
\end{figure} 

\sarif is a standard developed under OASIS~\cite{sarifSASP}. The technical committee of \sarif includes members from several static analysis tool vendors, including GrammaTech and other large-scale users~\cite{sarifSASP}. \sarif is a JSON-based format designed to not only report the results of an analysis but also its metadata, including schema, URI, and version. It has been created with the goal of unifying the output format of different static analysis tools, making it easy to integrate the reports into a single interface, which is the main objective of Static Analysis Server Protocol (\sasp)~\cite{sarifSASP}.

\sasp acts as a service where clients, such as the Eclipse Integrated Development Environment (IDE)\footnote{https://www.eclipse.org/ide/}, IntelliJ IDEA\footnote{https://www.jetbrains.com/idea/}, or Visual Studio Code\footnote{https://code.visualstudio.com/} can request static analysis results obtained from other analysis tools for a given program to analyze, as illustrated in Figure~\ref{fig:sasp}. For such a service to respond to a query quickly, it is necessary to enforce a common output standard to aggregate all analysis warnings results efficiently. \sasp achieves this by leveraging \sarif.

We explore how to make an analysis tool support \sarif, in order to eventually incorporate it in the \sasp system, thus enabling interoperability and potential integration with other static analysis tools. In particular, we focus on \cognicrypt~\cite{cogniCrypt}, a static analysis tool that detects misuses of cryptographic APIs in Java programs. The current version of \cognicrypt returns its results in its own format, which is used to display warning traces in Eclipse. \cognicrypt is implemented as an Eclipse plugin, and provides software developers with two main functionalities:
\begin{itemize}
	\item generating secure implementations of common cryptographic programming tasks,
	\item and analyzing developer code in the IDE and reporting existing misuses of cryptographic libraries.
\end{itemize}

In this paper, we first present \cognicrypt's original reporting format in Section~\ref{cognicrypt}. We then detail the \sarif format and explain its structure and syntax in Section~\ref{sarif}. Then, Section~\ref{approach} describes our implementation of the connector that exports \cognicrypt results in \sarif format. Finally, Section~\ref{summary} summarizes the outcomes of this paper and presents future work.

\section{The \cognicrypt Report Format}
\label{cognicrypt}

\begin{figure}[t]
\centering
\begin{lstlisting}[caption={Code example containing a ConstraintError and a TypestateError}, label={list:example}, language=Java, escapechar=|]
public class Crypto {
	public void getKey(int keySize) throws NoSuchAlgorithmException{
		KeyGenerator c = KeyGenerator.getInstance("AES");
		if (keySize > 0)
			c.init(512);
		else
			c.generateKey(); |\label{line:tse}|
		c.generateKey();
	}
}
\end{lstlisting}
\end{figure}

Cryptography is used for many different purposes. From hashing to encrypting, complex cryptographic libraries are used in many applications. However, using those libraries is not straightforward. Recent studies indicate that software developers have limited to no knowledge on the usage of APIs of cryptographic libraries. Lazar et al.~\cite{caseStudyProblems} carried out an investigation on 269 cryptography related vulnerabilities and found that 83\% of them resulted from software application developers misusing the cryptographic libraries. Nadi et al.~\cite{jumpHoops} show that most cryptographic misuses are due to the insufficient knowledge on the library usage by the developer, and that developers require debugging tools in their development environments to support them.

In order to detect cryptographic API misuses, \cognicrypt uses a set of cryptographic rules encoded in the \crysl format, a definition language that allows cryptographic experts to encode the secure usage of cryptographic libraries in a light-weighted syntax. \cognicrypt automatically converts those rules into an efficient flow-sensitive and context-sensitive static data-flow analysis that it then runs to detect the API misuses described by the rules. In its current state, \cognicrypt contains a complete ruleset for the APIs of the Java Cryptography Architecture (JCA).

\begin{figure}[t]
\centering
\begin{lstlisting}[caption={\cognicrypt console output for Listing~\ref{list:example}}, label={list:outputexample}, language=Java]
Findings in Java Class: Example.Crypto

	 in Method: void getKey(int)
		ConstraintError violating CrySL rule for javax.crypto.KeyGenerator (on Object #bfd7ff31836bf8643830e32ce26e9ef95 4d0522793ed0e9722ce44f0b255d4ef)
			First parameter (with value 512) should be any of {128, 192, 256}
			at statement: virtualinvoke r1.<javax.crypto.KeyGenerator: void init(int)>(varReplacer29)
			at line: 5

		TypestateError violating CrySL rule for javax.crypto.KeyGenerator (on Object #bfd7ff31836bf8643830e32ce26e9ef95 4d0522793ed0e9722ce44f0b255d4ef)
			Unexpected call to method generateKey on object of type javax.crypto.KeyGenerator.
			at statement: virtualinvoke r1.<javax.crypto.KeyGenerator: javax.crypto.SecretKey generateKey()>()
			at line: 7
\end{lstlisting}
\end{figure}

In \cognicrypt, each \crysl rule defines the correct use of a specific Java class of a cryptography library, by encoding constraints on usage order of API calls and parameter types. Error types and reporting are also encoded in \crysl.
When \cognicrypt analyses a Java program, a listener waits for the generation of analysis results and outputs them in the command-line as they are returned. A developer can change the reporting format by implementing their own custom reporting listener and using it in place of the default command-line listener. \cognicrypt supports seven types of errors:

\begin{itemize}[leftmargin=.5cm]
\item \textbf{ConstraintError:} This type of error refers to the wrong parameters being supplied to particular method calls. For example, calling \texttt{Cipher.getInstance("AES")} instead of the secure version \texttt{Cipher.getInstance("AES/ECB/PKCS5Padding")}.

\item \textbf{NeverTypeOfError:} This error is reported when a variable is of an insecure type, such as a password contained in a string instead of a char array.

\item \textbf{ForbiddenMethodError:} This error is raised when a deprecated or insecure method is called, such as the constructor \texttt{PBEKeySpec(char[] password)}.

\item \textbf{TypestateError:} When a call to a method is issued when it shouldn't be, \cognicrypt raises a TypestateError. For example, calling \texttt{Cipher.doFinal()} when no call to \texttt{Cipher.init()} has been issued before.

\item \textbf{RequiredPredicateError:} This error refers to a second-degree ConstraintError: when an object requires another object to be used in a specific way, and this was not the case. For example, a \texttt{Cipher} object receiving a hardcoded key will raise an error, since such keys should not be hardcoded.

\item \textbf{ImpreciseValueExtractionError:} This error is used when the analysis could not retrieve the parameter passed to a cryptographic method, for example when a key size is supplied in a configuration file instead of in the code. Since the parameter could be faulty, an error of lesser importance is raised.

\item \textbf{IncompleteOperationError:} This error relates to the TypestateError, but instead of referring to a wrong method call, it is raised when a missing call is detected. An example is never calling \texttt{Cipher.doFinal()} on a cipher object.
\end{itemize}

We illustrate a ConstraintError and a TypestateError in Listing~\ref{list:example}, with \cognicrypt's corresponding report shown in Listing~\ref{list:outputexample}.
Listing~\ref{list:example} presents a Java method which generates a cryptographic key using an instance of \texttt{KeyGenerator}. Two errors are made here: first, \texttt{init()} of \texttt{KeyGenerator} is called using an incorrect parameter: \texttt{512} instead of the secure \texttt{128}, \texttt{192}, or \texttt{256} values. Second, along the \emph{else} path, the key generator object is never initialized before \texttt{generateKey()} is called.
Using the \crysl rules that describe the usage of \texttt{KeyGenerator}, \cognicrypt thus detects the two errors as a ConstraintError and a TypestateError. We show the corresponding \crysl rules in Appendix~\ref{appendix:key_generator}.

When reporting an error, \cognicrypt provides:
\begin{itemize}
\item The error type.
\item The error location, as a line number and file name.
\item A customized error message. For example, for the ConstraintError in Listing~\ref{list:outputexample}, the error message contains the erroneous first parameter of \texttt{getKey()}, and provides other parameters that should be used instead.
\end{itemize}



\section{The \sarif Format}
\label{sarif}

We now detail the \sarif specification, with respect to reporting warnings. The complete \sarif documentation is found online~\cite{sarifSASP}.



\sarif is a JSON format standard~\cite{sarifDoc}. Its three main root keys--shown in Listing~\ref{list:mainpairs}--are: \emph{version} which specifies the version of the \sarif format, \emph{\$schema} which specifies the URI of the predefined JSON schema corresponding to the version, and \emph{runs} an array containing the results of the analysis runs. The six main subkeys of an individual run are shown in Listing~\ref{list:runs}. 

\begin{figure}[t]
\begin{lstlisting}[language=json, caption={Root key-value pairs of \sarif}, label={list:mainpairs}]
{
	"version": "2.0.0",
	"$schema": "http://json.schemastore.org/sarif-2.0.0",
	"runs": [{..}]
}
\end{lstlisting}
\end{figure}

\begin{figure}[t]
\begin{lstlisting}[language=json, caption={Subkeys of the key \textit{runs} in \sarif}, label={list:runs}]
"runs": [{
	"tool": {..},
	"invocations": [{..}],
	"files": {..},
	"logicalLocations": {..},
	"results": [{..}],
	"resources": {..}
}]
\end{lstlisting}
\end{figure}

\begin{figure}[t]
\begin{lstlisting}[language=json, caption={Subkeys of the key \textit{invocations} in \sarif}, label={list:invocation}, escapechar=|]
"invocations": {
	"commandLine": "java -cp CryptoAnalysis-1.0.0-jar-with-	dependencies.jar crypto.HeadlessCryptoScanner --rulesDir=src/test/resources/ --applicationCp=CogniCryptDemoExample/ Examples.jar  --sarifReport --reportDir=CogniCrypt/reports",|\label{line:sarifInvocation}|

	"responseFiles": [{
          "uri": "CryptoAnalysis/build/ CryptoAnalysis-jar-with-dependencies.jar",
          }, {
          "uri": "CryptoAnalysis/src/test/resources/",
          }, {
          "uri": "CryptoAnalysisTargets/ CogniCryptDemoExample/ Examples.jar"
          }],
	"startTime": "2016-07-16T14:18:25Z",
	"endTime": "2016-07-16T14:19:01Z",
	"fileName": "CryptoAnalysis/build/ CryptoAnalysis-jar-with-dependencies.jar",
	"workingDirectory": "/home/CryptoAnalysis/",
	"environmentVariables": {
		"PATH": "..",
		"HOME": "..",
	},
	"configurationNotifications": [{
		"level": "error",
		"message": {
			"text": "ERROR StatusLogger No Log4j 2 configuration file found. Using default configuration (logging only errors to the console)."
		}
	}],
	"toolNotifications": [{
		"level": "note",
		"message": {
			"text": "Finished initializing soot."
		}
	}, {
		"level": "warning",
		"message": {
			"text": "Couldn't find any method for CryptSLMethod: keyMaterial = javax.crypto.SecretKey. getEncoded();"
		},
	}, {
		"level": "note",
		"message": {
			"text": "Static Analysis took 1 seconds!."
		}
	}]
}
\end{lstlisting}
\end{figure}

The syntax of the \emph{runs} key can be separated into two categories:
\begin{itemize}
    \item reporting analysis results (\emph{invocations}, \emph{files}, \emph{results}, and \emph{logicalLocations} keys), which we detail in Section~\ref{results},
    \item analysis metadata (\emph{tool} and \emph{resource} keys), which we explore in Section~\ref{metadata}.
\end{itemize}

\subsection{Reporting Analysis Results}\label{results}

In this section, we detail the \emph{invocations}, \emph{files}, \emph{results}, and \emph{logicalLocations} keys and their subkeys.

\paragraph{invocations}
The \emph{invocations} key describes the invocation information of the static analysis tool that was run. Invocation information mainly includes the start time of the analysis, the end time of the analysis, the environmental variables that are used to run the analysis, the command that is used to invoke the analysis, and the notifications displayed during the analysis. 
Those notifications are categorized into configuration notifications and tool notifications. The former contain notification objects describing the conditions relevant to the tool configuration, while the latter describe the runtime environment after the static analysis is invoked. A snippet of a \cognicrypt invocation object is shown in Listing~\ref{list:invocation}.

\paragraph{files}
The \emph{files} key contains the information of all the files relevant to the run: the files in which analysis results were detected, or all files examined by the analysis tool. In some cases, a file might be nested inside another file (for example, in a compressed container), which is then referred to as its \emph{parent}. 
In the case of nested files, the parent's name is separated from nested fragment with the character, ``\texttt{\#}''. The nested fragment then starts with  ``\texttt{/}''. An example where the file ``intro.docx'' is located in the file ``app.zip'' is shown in Listing~\ref{list:files}.

\begin{figure}[t]
\begin{lstlisting}[language=json, caption={Subkeys of the key \textit{files} in \sarif}, label={list:files}]
"files": {
	"collections/list.cpp": {
		"mimeType": "text/x-c",
		"length": 980,
	},
	"app.zip#/docs/intro.docx": {
		"uri": "/docs/intro.docx",
		"mimeType":"wordprocessingml.document",
	"parentKey": "app.zip",
	"length": 4050
	}
}
\end{lstlisting}
\end{figure}

\paragraph{logicalLocations}
The optional key \emph{logicalLocations} is used in case the analysis tool yields results that include physical location information, (e.g., source file name, the line and column numbers) and logical location information (e.g., namespace, type, and method name). 
In some cases, a logical location might be nested in another logical location referred to as its \emph{parent}. In such cases, \textit{logicalLocations} should contain properties describing each of its parents, up to the top-level logical location. An example of a warning  detected in the C++ class \textit{namespaceA::namespaceB::classC} is shown in Listing~\ref{list:logical}. The corresponding \emph{logicalLocations} object contains the properties describing the class along with its containing namespaces.


\begin{figure}[t]
\begin{lstlisting}[language=json, caption={Subkeys of the key \textit{logicalLocations} in \sarif}, label={list:logical}]
"logicalLocations": {
	"namespaceA::namespaceB::classC": {
		"name": "classC",
		"kind": "type",
		"parentKey": "namespaceA::namespaceB"
	},
	"namespaceA::namespaceB": {
		"name": "namespaceB",
		"kind": "namespace"
		"parentKey": "namespaceA"
	},
	"namespaceA": {
		"name": "namespaceA",
		"kind": "namespace"
	}
}
\end{lstlisting}
\end{figure}

\begin{figure}[t]
\begin{lstlisting}[language=json, caption={Subkeys of the key \textit{results} in \sarif}, label={list:results}]
"results": [{
	"ruleId": "C2001",
	"ruleMessageId": "default",
	"richMessageId": "richText",
	"message": {
		"text": "Deleting member 'x' of variable 'y' may compromise performance on subsequent accesses of 'y'."
	},
	"suppressionStates": [ "suppressedExternally" ],
	"baselineState": "existing",
	"level": "error",
	"analysisTarget": {
		"uri": "collections/list.cpp",
	},
	"locations": [{..}],
	"codeFlows": [{..}],
	"stacks": [{..}],
	"fixes": [{..}],
	"workItemUris": [
		"https://github.com/example/project/issues/42",
		"https://github.com/example/project/issues/54"
	]
}]
\end{lstlisting}
\end{figure}

\begin{figure}[t]
\begin{lstlisting}[language=json, caption={Subkeys of the key \textit{locations} in \sarif}, label={list:loc}]
"locations":[{
	"physicalLocation": {
		"fileLocation": {
			"uri": "collections/list.h",
		},
	"region": {
		"startLine": 15,
		"startColumn": 9,
		"endLine": 15,
		"endColumn": 10,
		"charLength": 1,
		"charOffset": 254,
		"snippet": {
			"text": "add_core(ptr, offset, val);\n    return;"
			}
		}
	},
	"fullyQualifiedLogicalName": "collections::list:add"
}]
\end{lstlisting}
\end{figure}

\paragraph{results}
Each \emph{run} object contains an array of result objects, under the key \emph{results}. Each result represents a warning reported by the analysis, an example of which is shown in Listing~\ref{list:results}. We now detail the subkeys of a run object.

\begin{figure}[t]
\begin{lstlisting}[language=json, caption={Subkeys of the key \textit{codeFlow} in \sarif}, label={list:codeFlow}]
"codeFlows": [{
    "message": {
      "text": "Path from declaration to usage"
    },
    "threadFlows": [
      {
        "id": "thread-52",
        "locations": [
          {
            "step": 1,
            "importance": "essential",
            "message": {
              "text": "Variable \"ptr\" declared.",
            },
            "location": {...
                "region": {
                  "startLine": 15,
                  "snippet": {
                    "text": "int *ptr;"
                  },
                }
              },
            },
            "module": "platform"
          },
          {
            "step": 2,
            "importance": "essential",
            "message": {
              "text": "Uninitialized variable \"ptr\" passed to
                       method \"add_core\".",
              "richText": "Uninitialized variable `ptr` passed to
                           method `add_core`."
            },
            "location": {
              "physicalLocation": {...
              	"region": {
                  "startLine": 25,
                  "snippet": {
                    "text": "add_core(ptr, offset, val)"
                  }
                }
              },
            },
          }
        ]
      }
    ]
  }
],
\end{lstlisting}
\end{figure}

\begin{figure}[t]
\begin{lstlisting}[language=json, caption={Subkeys of the key \textit{stacks} in \sarif}, label={list:stacks}]
"stacks": [{
    "message": {
      "text": "Call stack resulting from usage of uninitialized variable."
    },
    "frames": [
      {
        "message": {
          "text": "Exception thrown."
        },
        "location": {
          "physicalLocation": {...
            "region": {
              "startLine": 110,
              "startColumn": 15
            }
          },
        },
        "threadId": 52,
        "address": 10092852,
        "parameters": [ "null", "0", "14" ]
      },
      {
        "location": {
          "physicalLocation": {...
            "region": {
              "startLine": 43,
              "startColumn": 15
            }
          },
        },
        "threadId": 52,
        "address": 10092176,
        "parameters": [ "14" ]
      },
      {
        "location": {
          "physicalLocation": {...
            "region": {
              "startLine": 28,
              "startColumn": 9
            }
          },
        },
        "threadId": 52,
        "address": 10091200,
      }
    ]
  }
],
\end{lstlisting}
\end{figure}

\begin{itemize}

\item \emph{ruleId} is the unique identifier of the analysis rule that was evaluated to produce the result.

\item \emph{ruleMessageId} refers to a message in the metadata.

\item \emph{richMessageId} refers to a more descriptive message in the metadata.

\item \emph{message} describes the warning. If the message is not specified, the \emph{ruleMessageId} is used instead.

\item \emph{baselineState} describes the state of the result with respect to a previous baseline run (i.e., new, existing, or absent).

\item \emph{level} indicates the severity of the result (e.g., error, warning).

\item \emph{locations} contains one or more unique location objects marking the exact location of warning, as shown in Listing~\ref{list:loc}. It contains the physical location (e.g., file name, line and column) or the logical location (such as namespace, type, and method name) and the region in the file where the result is found. If the physical location information is absent, the \textit{fullyQualifiedLogicalName} property is used instead. 

\item \emph{codeFlows} is an array of individual code flows, which describe the execution path of the warning step by step. An example is shown in Listing~\ref{list:codeFlow}.

\item \emph{stacks} is an array of call-stack frames created by the analysis tool. Each stack frame contains location information to the call-stack object, a thread id, parameter values, memory addresses, etc. This is illustrated in Listing~\ref{list:stacks}.

\item \emph{fixes} is an array of fix suggestions. For each file in a fix object, the format describes regions that can be removed and new contents to be added. An example is found in Listing~\ref{list:fixes}.

\item \emph{workItemUris} is an array of URIs to existing work items associated with the warning. Work items can be GitHub issues or JIRA tickets for example.

\end{itemize}

\begin{figure}[t]
\begin{lstlisting}[language=json, caption={Subkeys of the key \textit{fixes} in \sarif}, label={list:fixes}]
"fixes": [{
	"description": {
		"text": "Initialize the variable to null"
		},
	"fileChanges": [{
		"fileLocation": {
			"uri": "collections/list.h",
			},
		"replacements": [{
			"deletedRegion": {
				"startLine": 42
			},
			"insertedContent": {
				"text": "A different line\n"
			}
		}]
	}]
}]
\end{lstlisting}
\end{figure}

\subsection{Metadata}\label{metadata}

We now detail the \emph{tool} and \emph{resources} keys and their subkeys, which are used in \sarif to store analysis metadata.

\paragraph{tool}
The key \emph{tool} contains information regarding the static analysis tool that performed the analysis and produced the report. Its self-descriptive keys are shown in Listing~\ref{list:tool}.

\paragraph{resources}
The \emph{resources} key contains resource objects such as localized items such as rule metadata and message strings associated with the rules. This prevents data duplication if, for example, multiple warnings refer to the same rule. Each rule object contains rule information such as rule id, rule description, and message strings. This is illustrated in Listing~\ref{list:resources}. Note that the subkeys \emph{messageStrings} and \emph{richMessageStrings} contain all of the \emph{messageStrings} and \emph{richMessageStrings} of the result objects (Listing~\ref{list:resources}).

\section{From the \cognicrypt Reporting Format to \sarif}
\label{approach}

In this section, we detail our approach for converting \cognicrypt results to the \sarif format, following the requirements of Section I.2 of the SARIF documentation\footnote{\url{http://docs.oasis-open.org/sarif/sarif/v2.0/csprd01/sarif-v2.0-csprd01.html\#_Toc517436281}}~\cite{sarifDoc}. To illustrate our implementation, we use the example \cognicrypt report in Listing~\ref{list:examplesoutput} obtained after analysing an example file from \cognicrypt: \emph{Examples.jar}\footnote{\url{https://github.com/CROSSINGTUD/CryptoAnalysis/blob/master/CryptoAnalysisTargets/CogniCryptDemoExample/Examples.jar}}. The listing contains two warnings: a ConstraintError (lines 297-299) and a TypestateError (lines 303-305). Listings~\ref{list:cognijson}--~\ref{list:cogniresultsjson} are snippets of the same report in \sarif format, with the latter describing the warnings, and the former containing all of the remaining data and metadata. 

\subsection{Mapping \cognicrypt Data to \sarif Keys}
\label{sec:identifyingImpInformation}

To write a \sarif exporter for \cognicrypt, it is important to first identify which information to export from the \cognicrypt error format. We detail this information in this section.




The first level of the \sarif JSON hierarchy contains the \emph{version} and \emph{\$schema} information. In our implementation, this data is populated based on the current \sarif version: \texttt{2.0.0} (Listing~\ref{list:cognijson} line 307), and its respective schema reference (Listing~\ref{list:cognijson} line 308). This information is hardcoded in our converter.

\begin{figure}[t]
\begin{lstlisting}[language=json, caption={Subkeys of the key \textit{tools} in \sarif}, label={list:tool}]
"tool": {
	"name": "CodeScanner",
	"fullName": "CodeScanner 1.1 for Unix (en-US)",
	"version": "2.1",
	"semanticVersion": "2.1.0",
	"language": "en-US",
	"properties": {
		"copyright": "Copyright (c) 2017 by Example Corporation.
		All rights reserved."
		}
},
\end{lstlisting}
\end{figure}

\begin{figure}[t]
\begin{lstlisting}[language=json, caption={Subkeys of the key \textit{resources} in \sarif}, label={list:resources}]
"resources": 
{
	"rules": {
		"C2001": {
			"id": "C2001",
			"shortDescription": {
				"text": "A variable was used without being initialized."
			},
			"fullDescription": {
				"text": "A variable was used without being initialized. This can result in runtime errors such as null reference exceptions."
			},
			"messageStrings": {
				"default": "Variable \"{0}\" was used without being initialized."
			},
			"richMessageStrings": {
				"richText": "Variable `{0}` was used without being initialized."
			}
		}
	}
}
\end{lstlisting}
\end{figure}

\begin{figure}[t]
\begin{lstlisting}[language=json, caption={Example \cognicrypt output}, label={list:examplesoutput},  escapechar=|]
Findings in Java Class: example.TypestateErrorExample |\label{line:commandLineReportFullyQualName}|

	 in Method: getPrivateKey|\label{line:commandLineMethod}|
		ConstraintError violating CrySL rule for KeyPairGenerator (on Object #9367df75558b10b537d558f11cb 7a523f082e7e256ab7ba827a36db283cf940e)|\label{line:commandLineMessageText}|
			First parameter (with value 1024) should be any of {2048, 4096}|\label{line:commandLineRichText}|
			at line: 29|\label{line:commandLineStartLine}|


	 in Method: main
		TypestateError violating CrySL rule for Signature (on Object #9c822ffdf2268ba2e0ff61f394b200 a7510d25a3d4a558ae811e624191c3583b)
			Unexpected call to method sign on object of type java.security.Signature. Expect a call to one of the following methods initSign,update
			at line: 24
\end{lstlisting}
\end{figure}

\begin{figure}[t]
\begin{lstlisting}[language=json, caption={\sarif output example for Listing~\ref{list:examplesoutput} (1/2)}, label={list:cognijson},  escapechar=|]
{
  "version": "2.0.0",
  "$schema": "http://json.schemastore.org/sarif-2.0.0",
  "runs": [{
    "tool": {
      "semanticVersion": "1.0.0",
      "fullName": "CogniCrypt (en-US)",
      "language": "en-US",
      "version": "1.0.0"
    },
    "files": {
      "example/TypestateErrorExample.java": {|\label{line:sarifReportFileName}|
        "mimeType": "text/java"|\label{line:sarifReporterMimeType}|
      }
    },
    "results": [...],
    "resources": {|\label{line:sarifReportResources}|
      "rules": {
        "TypestateError": {
          "id": "TypestateError",
          "fullDescription": {
            "text": "The ORDER block of CrySL is violated, i.e., the expected method sequence call to be made is incorrect. For example, a Signature object expects a call to initSign(key) prior to update(data)."
          }
        },
        "ConstraintError": {
          "id": "ConstraintError",
          "fullDescription": {
            "text": "A constraint of a CrySL rule is violated, e.g., a key is generated with the wrong key size."
          }
        },
      }
    },
  }]
}
\end{lstlisting}
\end{figure}

To fill the \emph{runs} information (Listing~\ref{list:cognijson} line 309), we map the following data found in the \cognicrypt error format to the keys of the \sarif format:
\begin{itemize}
	\item \emph{tool} contains static information on \cognicrypt. The information to this key, such as \emph{semanticVersion}, \emph{version}, \emph{fullName}, and \emph{language} (Listing~\ref{list:cognijson} lines 311-314) is fetched from a hardcoded configuration file in our converter: \texttt{SARIFConfig}\footnote{\url{https://github.com/CROSSINGTUD/CryptoAnalysis/blob/develop/CryptoAnalysis/src/main/java/crypto/reporting/SARIFConfig.java}}. 
	\item \emph{files} data is available from the existing \cognicrypt reporter: \texttt{CommandLineReporter}\footnote{\url{https://github.com/CROSSINGTUD/CryptoAnalysis/blob/master/CryptoAnalysis/src/main/java/crypto/reporting/CommandLineReporter.java}}. The value for the key is fetched from the fully qualified name reported in line~\ref{line:commandLineReportFullyQualName}, and can be seen in line~\ref{line:sarifReportFileName}. Since \cognicrypt only supports Java projects, \emph{mimeType} is hardcoded to a default value: \texttt{text/java}, as shown in line~\ref{line:sarifReporterMimeType}.
	\item \emph{results} is populated from the information available from the current \cognicrypt report format. An example is shown in Listing~\ref{list:cogniresultsjson}. For the ConstraintError reported in line~\ref{line:commandLineMessageText}, \emph{locations.physicalLocation.fileLocation.uri} and \emph{fullyQualifiedName} are fetched from lines~\ref{line:commandLineReportFullyQualName} and~\ref{line:commandLineMethod}; \emph{locations.physicalLocation.fileLocation.region.startLine} is obtained from line~\ref{line:commandLineStartLine}; \emph{ruleId}, \emph{message.text} and \emph{message.richText} are populated from lines~\ref{line:commandLineMessageText} and~\ref{line:commandLineRichText}.
	\item \emph{resources} are illustrated shown in Listing~\ref{list:cognijson} from line~\ref{line:sarifReportResources}. The different types of rules reported by \cognicrypt are located in the package \texttt{crypto.analysis.errors}\footnote{\url{https://github.com/CROSSINGTUD/CryptoAnalysis/tree/master/CryptoAnalysis/src/main/java/crypto/analysis/errors}}. For each of those errors, a \emph{fullDescription} is retrieved from the configuration file \texttt{SARIFConfig}\footnote{\url{https://github.com/CROSSINGTUD/CryptoAnalysis/blob/develop/CryptoAnalysis/src/main/java/crypto/reporting/SARIFConfig.java}}.
	\item \emph{logicalLocations} are not available in \cognicrypt. Currently, \cognicrypt only reports an error at the line where it was detected and not the complete witness path of the warning. The information could be generated in the class \texttt{ErrorMarkerListener}\footnote{\url{https://github.com/CROSSINGTUD/CryptoAnalysis/blob/develop/CryptoAnalysis/src/main/java/crypto/reporting/ErrorMarkerListener.java}}, and then, exported in \sarif.
	\item \emph{invocations} is not implemented in our connector, but can be retrieved from CogniCrypt. The \emph{commandLine}, \emph{workingDirectory}, \emph{startTime}, and \emph{endTime} are stored in the class \texttt{HeadlessCryptoScanner}\footnote{\url{https://github.com/CROSSINGTUD/CryptoAnalysis/blob/develop/CryptoAnalysis/src/main/java/crypto/HeadlessCryptoScanner.java}}, and the \emph{toolNotifications} are generated in the class \texttt{CommandLineReporter}\footnote{\url{https://github.com/CROSSINGTUD/CryptoAnalysis/blob/develop/CryptoAnalysis/src/main/java/crypto/reporting/CommandLineReporter.java}}.
\end{itemize}

\begin{figure}[t]
\begin{lstlisting}[language=json, caption={\sarif output example for Listing~\ref{list:examplesoutput} (2/2)}, label={list:cogniresultsjson}]

	"results": [
	{
		"locations": [{
			"physicalLocation": {
				"fileLocation": {
				"uri": "example/TypestateErrorExample.java"
				},
				"region": {
					"startLine": 29
				}
			},
			"fullyQualifiedLogicalName": "example::TypestateErrorExample:: getPrivateKey"
		}],
		"ruleId": "ConstraintError",
		"message": {
		"text": "First parameter (with value 1024) should be any of {2048, 4096}.",
		"richText": "ConstraintError violating CrySL rule for KeyPairGenerator."
		}
	}, {
	"	locations": [{
			"physicalLocation": {
				"fileLocation": {
				"uri": "example/TypestateErrorExample.java"
				},
				"region": {
					"startLine": 24
				}
			},
			"fullyQualifiedLogicalName": "example::TypestateErrorExample::main"
			}],
			"ruleId": "TypestateError",
			"message": {
			"text": "Unexpected call to method sign on object of type java.security.Signature. Expect a call to one of the following methods initSign,update.",
			"richText": "TypestateError violating CrySL rule for Signature."
			}
		}]

	\end{lstlisting}
\end{figure}

\subsection{Implementation Details}

Our implementation of the \cognicrypt--\sarif converter is integrated in the \cognicrypt repository~\cite{sarifconnector} and can be enabled by using the \texttt{--sarifReport} option and specifying a directory to store the generated report using \texttt{--reportDir} option. An example is shown at line~\ref{line:sarifInvocation} of Listing~\ref{list:invocation}.

The results of \cognicrypt are available through the class \texttt{crypto.reporting.ErrorMarkerListener}. Each object of this class contains an \texttt{errorMarkers} field containing all warnings. The main class of our converter is \texttt{crypto.reporting.SARIFReporter}, which extends \texttt{ErrorMarkerListener}. In this class, we have overridden the method \texttt{afterAnalysis()}, in which we iterate through the \cognicrypt warnings and convert them into \sarif. Since \cognicrypt stores its results in a Google Guava Table, the complexity of our connector is linear with respect to the number of findings. 



\subsection{Evaluation}

We verified the implementation of our \cognicrypt converter using an online \sarif validator\footnote{\url{http://sarifweb.azurewebsites.net}}~\cite{sarifValidator}. The validator takes the generated \sarif file as the input, scans over it, and communicates the format issues when the generated \sarif report does not follow the standard specified in~\cite{sarifDoc}. We generated the \sarif files for all of the \cognicrypt test cases\footnote{https://github.com/CROSSINGTUD/CryptoAnalysis}, including the one used in this report (Listings~\ref{list:cognijson} and~\ref{list:cogniresultsjson}). The validation of the \sarif format passed. 

A threat to validity is that the validator was in beta-testing phase at the time. Thus, in addition, we manually verified the JSON format of our reports according to the \sarif standard. All of our \sarif reports were correct.


\section{Conclusion and Future Work}
\label{summary}

In this paper, we explored how to convert the \cognicrypt error format into the more general \sarif format. After detailing the two formats, we detailed our implementation. In our evaluation, we confirmed the correctness of our converter on the \cognicrypt test cases.
The current implementation of our connector is available online as part of the official \cognicrypt implementation on GitHub \cite{sarifconnector}. Since this is an initial prototype, there is still room for improvement. One such improvements is to finish the implementation of the converter to include invocation and logical location information. Another improvement concerns the \cognicrypt error format, which does not encode as many details as it could. For example call-graph information is available in the analysis and could be encoded in \sarif, but the data is lost through the \cognicrypt report. The connector can be improved to retrieve the information directly from the analysis. As a follow-up to this work, \cognicrypt also needs a full support for \sasp, since it is now able to export its results in \sarif. 

\begin{acks}
This research was conducted under the supervision of Eric Bodden as part of the Secure Systems Engineering seminar at Paderborn University, organized by Lisa Nguyen Quang Do. It was partially funded by the Heinz Nixdorf Foundation and by the NRW Research Training Group on Human Centered Systems Security (nerd.nrw).
\end{acks}

\bibliographystyle{ACM-Reference-Format}
\bibliography{paper}

\newpage

\appendix
\section{CrySL Rules}
\subsection{KeyGenerator}
\label{appendix:key_generator}
\begin{lstlisting}[language=Java]
SPEC javax.crypto.KeyGenerator
OBJECTS
    int keySize;
    java.security.spec.AlgorithmParameterSpec params;
    javax.crypto.SecretKey key;
    java.lang.String alg;
    java.security.SecureRandom ranGen;

EVENTS
    g1: getInstance(alg);
    g2: getInstance(alg, _);
    Gets := g1 | g2;

    i1: init(keySize);
    i2: init(keySize, ranGen);
    i3: init(params);
    i4: init(params, ranGen);
    i5: init(ranGen);
    Inits := i1 | i2 | i3 | i4 | i5;
    
    gk: key = generateKey();

ORDER
    Gets, Inits?, gk

CONSTRAINTS
	alg in {"AES", "HmacSHA224", "HmacSHA256", "HmacSHA384", "HmacSHA512"};
    alg in {"AES"} => keySize in {128, 192, 256};
   
REQUIRES
    randomized[ranGen];
    
ENSURES 
    generatedKey[key, alg];
\end{lstlisting}

\end{document}